\input{aipcheck}

\documentclass[
    ,final            
  ]
  {aipproc}

\layoutstyle{6x9}

\newcommand{\ms}{\mbox{m\,s$^{-1}$}}
\newcommand{\kms}{\mbox{km\,s$^{-1}$}}
\newcommand{\Msun}{\mbox{M$_{\odot}$}}

\newcommand{\gtsimeq}{\raisebox{-0.6ex}{$\,\stackrel
        {\raisebox{-.2ex}{$\textstyle >$}}{\sim}\,$}}

\begin{document}

\title{The Pan-Pacific Planet Search: A southern hemisphere search for 
planets orbiting evolved massive stars}


\author{Robert A.~Wittenmyer}{
  address={Department of Astrophysics, School of Physics, 
University of NSW, 2052, Australia}
}

\author{John Johnson}{
  address={Department of Astrophysics, California Institute of 
Technology, MC 249-17, Pasadena, CA 91125, USA}
}

\author{Liang Wang}{
  address={National Astronomical Observatories, Chinese Academy of 
Sciences, A20 Datun Road, Chaoyang District, Beijing 100012, China}
}

\author{Michael Endl}{
  address={McDonald Observatory, University of Texas at Austin, Austin, 
TX 78712, USA}
}

\begin{abstract}

The vast majority of known extrasolar planets orbit stars with a narrow 
range of masses (0.7-1.3 \Msun).  Recent years have seen rapid growth in 
our knowledge about the properties of planetary systems with host stars 
significantly more massive than the Sun.  Planet formation models 
predict that giant planets are more common around higher-mass stars 
($M_*>1.5$\Msun). However, these types of stars pose severe 
observational challenges while on the main sequence, resulting in a 
strong bias against them in current planet searches.  Fortunately, it is 
possible to obtain high-precision Doppler velocities for these massive 
stars as they evolve off the main sequence and cool as subgiants.  We 
describe the Pan-Pacific Planet Search, a survey of 170 subgiant stars 
using the 3.9m Australian Astronomical Telescope.  In collaboration with 
J.~Johnson's Keck survey of Northern ``retired A stars,'' we are 
monitoring nearly every subgiant brighter than $V=8$.  This survey will 
provide critical statistics on the frequency and characteristics of 
planetary systems formed around higher-mass stars.

\end{abstract}

\maketitle


\section{Introduction}

Radial-velocity planet searches have discovered more than 500 planets 
orbiting nearby stars.  The planets found to date span three orders of 
magnitude in mass, from tens of Jupiter masses down to a few Earth 
masses (Vogt et al.~2010, Mayor et al.~2009).  The known extrasolar 
planets also range in orbital separation from a few stellar radii to 
more than 5~AU, and encompass the full range of orbital eccentricities 
from circular ($e=0$) to incredibly elliptical ($e=0.94$).

The range of host-star masses probed by current radial velocity surveys, 
however, remains very narrow.  Most of the target stars fall in the 
range 0.7-1.3 \Msun (Johnson 2007, Valenti \& Fischer 2005).  This is a 
consequence of the technical requirements of Doppler exoplanetary 
detection, which demand that stars be cool enough to present an 
abundance of spectral lines, and rotate slowly enough that their 
absorption lines are not significantly broadened by rotation.  Stars of 
lower mass (e.g.~M dwarfs) are intrinsically faint in the optical, 
making the acquisition of high signal-to-noise spectra extremely 
expensive in telescope time (Endl et al.~2006).  Main sequence stars of 
higher mass have few usable absorption lines (due to their high 
temperatures), and also tend to be fast rotators ($v\sin~i>$ 50 \kms; 
Galland et al.~2005) due to their youth.  Only the most massive planets 
can be detected orbiting A and F dwarfs.  It is only recently that a 
significant number of planetary systems have been discovered orbiting 
massive stars.  These stars have proven to be a fertile hunting ground 
for interesting planetary systems, such as the 4:3 mean-motion resonant 
planets orbiting HD~200964 (Johnson et al.~2011).  Now, some headway is 
beginning to be made in addressing the crucial question of how planet 
formation depends on stellar mass (e.g.~Bowler et al.~2010, Johnson et 
al.~2010, Sato et al.~2010).

\subsection{The critical role of subgiants}

In the core-accretion theory of planet formation (Lissauer 1995, Pollack 
et al.~1996), solids in the protoplanetary disk accrete into rocky 
cores.  When these cores reach about 10\,$M_{\rm earth}$, they have 
sufficient gravity to rapidly accumulate disk gas and form giant 
planets, with final masses $M_p \gtsimeq 100\,M_{\rm earth}$.  A 
prediction of this model is that planet mass should positively correlate 
with host star mass, since increasing the mass of the star also 
increases the mass of the protoplanetary disk. Giant planets should then 
be more common around higher-mass stars.  Observational results to date 
support this prediction (Johnson et al.~2010, Johnson 2007).  The number 
statistics are poor at both the high- and low-mass ends of this trend, 
but suggest that high-mass stars are indeed more likely to host Jovian 
planets.  Results from the Lick and Keck survey of evolved A stars 
(Johnson et al.~2010) indicate a planet occurrence rate approaching 20\% 
for stars with $M_*>$1.5\Msun.  Further headway has been made at the 
low-mass end of this trend: Endl et al.~(2006) derive an upper limit for 
the giant planet occurence rate of 1.3\% in a sample of 90 M dwarfs, 
while Johnson et al.~(2007b) derive a similar rate (1.8$\pm$1.0\%) from 
their Keck M dwarf survey which was sensitive to lower masses and longer 
periods.

The core-accretion model also predicts that metal-rich disks are more 
efficient at forming cores, due to their enhanced surface density of 
solids.  Detailed planet-formation models by Ida \& Lin (2004) predicted 
that metal-rich protostellar disks would be more likely to form 
detectable planets.  Gonzalez (1999) first noted that planet-hosting 
stars appeared to be unusually metal-rich.  The current sample of 
extrasolar planetary systems appears to support this prediction: Fischer 
\& Valenti (2005) show that stars with twice solar metallicity 
([Fe/H]=+0.3) are about 3 times more likely to host a planet than stars 
with solar metallicity.

However, the causal mechanism for this observed relationship is still 
not completely settled.  While the core-accretion scenario does predict 
that metal-rich stars will preferentially form planets, an alternate 
``pollution'' model can also explain this effect.  In this scenario, 
planet host stars are metal-rich because debris from the planets' 
formation has enriched the surface layers of the star.  In this case, 
the enhanced metallicity of the star extends only to the convective zone 
(Laughlin \& Adams 1997).  These two hypotheses can be tested by 
determining the planet-metallicity relationship for \textit{subgiant} 
stars.  When a star evolves off the main sequence, the convective zone 
increases in size by about a factor of 35 (Pasquini et al.~2007).  If 
the high metallicities observed in planet hosts are due to pollution, 
this expansion of the convective zone will significantly dilute the 
extent of that pollution, and the subgiant's photosphere would return to 
its ``birth'' metallicity. Hence, one would \textit{not} expect a 
significant correlation between metallicity and planet frequency for 
subgiants.  If the enhanced metallicity of planet hosts is primordial in 
origin, however, then the planet-metallicity correlation observed for 
dwarf stars would also hold for subgiants.  Previous investigations into 
this issue have so far shown no evidence for the pollution scenario 
(Valenti \& Fischer 2008).  \citet{quirrenbach} have recently presented 
tentative evidence for a planet-metallicity correlation in their sample 
of 373 K giants.  

Subgiants (i.e. high-mass stars that have just evolved off the main 
sequence) provide a means to address two critical questions facing 
exoplanetary science.  First, they develop the multitude of narrow 
absorption lines critical for precision radial velocity work, because 
the photospheres of these stars expand and cool as they leave the main 
sequence, their rotational velocities drop.  Second, their ``birth'' 
metallicity is revealed as any potential pollution effects are diluted 
when their convective zones expand.

We have begun a survey of Southern metal-rich ([Fe/H]$>$0.0) subgiants 
which will tackle both these key issues -- planet frequency for 
``retired A stars,'' and planet frequency as a function of birth 
metallicity in the same stars.  Our goal is to enlarge the growing 
sample of planets around massive stars to strengthen the emerging trend 
between stellar mass and planet occurrence.  We will also investigate 
the planet-metallicity correlation among evolved stars in order to test 
pollution scenarios.  Our completed survey will dominate the number 
statistics for exoplanetary detections around metal-rich subgiants.

%
%

\section{The Pan-Pacific Planet Search}

\subsection{Target Selection}

This program is using the 3.9m Anglo-Australian Telescope (AAT) 
to observe a metal-rich sample of Southern Hemisphere subgiants.  We 
have selected 170 Southern stars with the following criteria: $1.0 < 
(B-V) < 1.2$, $1.8 < M_V < 3.0$, and $V<8.0$.  By requiring $(B-V)>1$, 
we extend the red limit of the Johnson et al.~(2006b) survey to the 
colours redward of $(B-V)=1.0$ that stellar models indicate will be 
dominated by metal-rich subgiants (Girardi et al.~2002).  This will 
allow us to obtain improved planetary detection statistics at 
[Fe/H]$>$0.0, and also (in light of the observed positive correlation 
between stellar metallicity and planet occurrence) deliver a roughly 
equivalent number of planetary detections to that obtained at Lick and 
Keck (though for metal-rich hosts).  At the same time, by requiring 
$M_V>1.8$, we exclude giant-branch stars, which have significant 
intrinsic velocity noise (``jitter'') due to stellar activity and 
pulsations (Saar et al.~1998, Wright 2005) -- typically about 20 \ms\ 
(Hekker et al.~2006).  Our target list includes about 30 stars from the 
Lick survey; this overlap will serve as a check on the systematics 
between the two telescopes.  These 170 metal-rich stars complement the 
metallicity-unbiased Northern stars in the Lick \& Keck survey.  
Together, the three telescopes are observing more than 600 stars: nearly 
every subgiant in the entire sky brighter than $V=8$.

\subsection{Progress to date}

This program has been awarded long-term status at the AAT, with 10 
guaranteed nights per semester through 2012 July.  As of 2010 October, 
we have obtained 3-5 observations for every target, with a 
signal-to-noise (S/N) of 100 per pixel.  Only 55\% of assigned time has 
resulted in usable data due to unusually poor weather conditions 
resulting from a prolonged La Ni\~na event in 2009-10.

As we are using the iodine-cell method to obtain precision Doppler 
velocities (Valenti et al.~1995, Butler et al.~1996), the spectra are 
superimposed with a forest of I$_2$ absorption lines between 5000 and 
6200 Angstroms.  These sharp features are used for a high-precision 
determination of the spectrograph point-spread function, which in turn 
enables us to correct for instrumental effects and obtain a precise 
radial velocity measurement.  We will obtain Doppler velocities using 
the \textit{Austral} code as first discussed in Endl et al.~(2000).  
\textit{Austral} is a proven Doppler code which has been used by the 
McDonald Observatory planet search programs for nearly 10 years 
(e.g.~Endl et al.~2004, 2006; Wittenmyer et al.~2009).

Nearly all spectra obtained to date contain iodine lines, but we can use 
the iodine-free regions of these spectra to perform preliminary stellar 
abundance analysis.  The equivalent widths (EWs) of over 200 lines from 
21 elements will be measured, including three light elements (C, N, O), 
four $\alpha$-elements (Mg, Si, Ca, Ti), three odd-Z light elements (Al, 
K, Sc), seven iron peak elements (Ti, V, Cr, Mn, Fe, Co, Ni), and five 
neutron-capture elements (Y, Ba, La, Pb, Eu).  The abundances are 
determined by the line synthesis program ABONTEST8 by P.~Magain (Liege, 
Belgium), based on model atmospheres interpolated by a plane-parallel, 
homogeneous and local thermodynamic equilibrium (LTE) model grid by 
Kurucz \& Bell (1995).  Non-LTE effects are taken into account for 
several elements (e.g.~O, Mg, Al).  Israelian et al.~(2009) reported 
that Lithium is depleted in planet-host dwarfs, while this phenomenon 
still remain unclear for higher mass stars.  A spectral synthesis will 
be performed for Li 6707 Angstroms to determine the Li abundances.  
Using the high resolution, high signal-to-noise spectra in this planet 
search program, the detailed chemical abundances, as well as basic 
stellar parameters of the subgiants can be obtained.  It is important to 
understand stellar nucleosynthesis and the chemical evolution history of 
the Galaxy by studying these stars in post-main-sequence stage.  These 
data will also help us to investigate different patterns of chemical 
abundance between subgiants with and without planets, and thus 
understand planet formation process around different types of stars.

%


\begin{theacknowledgments}

RW acknowledges support from a UNSW Vice-Chancellor's Fellowship.  We 
are grateful to the AAT Time Allocation Committee for granting this 
program long-term status.

\end{theacknowledgments}



\bibliographystyle{aipproc}   

\bibliography{sample}

\IfFileExists{\jobname.bbl}{}
 {\typeout{}
  \typeout{******************************************}
  \typeout{** Please run "bibtex \jobname" to optain}
  \typeout{** the bibliography and then re-run LaTeX}
  \typeout{** twice to fix the references!}
  \typeout{******************************************}
  \typeout{}
 }

\end{document}